\documentclass[twocolumn]{aastex631}

\newcommand{\dg}{^\circ}

\shorttitle{Eccentric neutron star disk driven type II outbursts}
\shortauthors{A. Franchini and R. G. Martin}
\graphicspath{{./}{newfigures/}}

\begin{document}

\title{Eccentric neutron star disk driven type II outburst pairs in Be/X-ray binaries}

\correspondingauthor{Alessia Franchini}
\email{alessia.franchini@unimib.it}

\author[0000-0002-8400-0969]{Alessia Franchini}
\affiliation{Dipartimento di Fisica ``G. Occhialini", Universit\'a degli Studi di Milano-Bicocca \\
Piazza della Scienza 3 \\
20126 Milano, Italy}

\author[0000-0003-2401-7168]{Rebecca G. Martin}
\affiliation{Department of Physics and Astronomy, University of Nevada \\
4505 South Maryland Parkway \\
Las Vegas, NV 89154, USA}



\begin{abstract}
Be star X-ray binaries are transient systems that show two different types of outbursts. Type I outbursts occur each orbital period while type II outbursts have a period and duration that are not related to any periodicity of the binary system. Type~II outbursts may be caused by mass transfer to the neutron star from a highly eccentric Be star disk. A sufficiently misaligned Be star decretion disk undergoes secular Von Zeipel-Lidov-Kozai (ZLK) oscillations of eccentricity and inclination.  Observations show that in some systems the type~II outbursts come in pairs with the second being of lower luminosity. We use numerical hydrodynamical simulations to explore the dynamics of the highly misaligned disk that forms around the neutron star as a consequence of mass transfer from the Be star disk. We show that the neutron star disk may also be ZLK unstable and that the eccentricity growth leads to an enhancement in the accretion rate onto the neutron star that lasts for several orbital periods, resembling a type II outburst. 
We suggest that in a type II outburst pair, the first outburst is caused by mass transfer from the eccentric Be star disk while the second and smaller outburst is caused by the eccentric neutron star disk. 
We find that the timescale between outbursts in a pair may be compatible with the observed estimates. 

\end{abstract}



\section{Introduction} 
\label{sec:intro}

Be star X-ray binaries are double systems composed of a neutron star that is accreting material from a companion Be star. The orbital periods are typically tens to hundreds of days.
Be stars are hot massive early type main sequence stars with a B spectral type (mass in the range $3-20\,M_{\odot}$) whose spectrum has shown one or more Balmer H-$\alpha$ emission lines \citep{Porter2003}. They are also rapidly rotating close to their break up velocity \citep{Slettebak1982}.
The stars brightness and spectra have been found to vary in time.  
Their emission is associated with the presence of circumstellar gas in the form of a relatively thin decretion disk \citep{Pringle1991,Porter2003}. \cite{Rajoelimanana2011} studied the properties of the optical emission from Be stars and found superorbital quasi-periodic variations on timescales of hundreds to thousands of days that are believed to be related to the formation and the depletion of the circumstellar disk.

Some of these systems have a transient nature owing to the interaction between the material from the Be star disk flowing onto and being accreted by the neutron star.
Be/X-ray transients typically show two types of outburst \citep{Stella1986, Negueruela1998}. Type~I outbursts are less luminous and occur each orbital period.
The occurrence of type~I outbursts can be explained either by the eccentric nature of the binary orbit that allows the neutron star to rip off material from the Be star disk close to periastron passage \citep{Okazaki2001,Negueruela2001,Okazaki2013}  
or an eccentric disk in systems with small binary eccentricity \citep{franchini2019b,martinfranchini2021}.

Giant or type II outbursts have luminosities close to the Eddington limit and no periodic modulation. In general, type~II outbursts  last several binary orbital periods \citep{Kretschmar2012}. \cite{Reig2018} found that type~II outbursts often occur in pairs with the second one being less luminous than the first.
Type II outbursts may be explained by the presence of a highly misaligned Be star disk that becomes eccentric due to the Von Zeipel-Lidov-Kozai (ZLK) mechanism  \citep{VonZeipel1910,Kozai1962,Lidov1962} and transfers mass through Roche lobe overflow onto the neutron star
\citep{martin2014,martin2014be,fu2015b,franchini2019a}. 
The ZLK oscillations and mass transfer timescales match the observed intervals between type II outbursts provided the binary orbital period is $\lesssim 150$ days and the Be star disk is flared \citep{martinfranchini2019}.

As material is transferred from the Be star disk to the neutron star, a compact disk forms around the neutron star \citep{Hayasaki2004,Hayasaki2006}. Because of the inclination of the Be star disk, the neutron star disk is likely misaligned and initially eccentric  \citep{franchini2019a}.
This idea is supported by the detection of quasi-periodic oscillations (QPOs) in some of the outbursts that are believed to be associated with the formation of an accretion disk around the neutron star \citep{Finger1996,Wilson2002}.

In this paper we extend these previous works to explore the dynamics of the accretion disk that forms around the neutron star.
In Section~\ref{sec:bedisk} we present the results of a simulation of a misaligned Be star disk. We  show that the disk can form around the neutron star with a sufficiently high inclination to undergo ZLK oscillations itself.  In Section~\ref{sec:nsdisk} we show a second simulation of the neutron star disk evolution with a higher resolution.
We find that the ZLK oscillation of the neutron star disk causes a long lasting accretion phase onto the compact object that resembles a type~II outburst.
We draw our conclusions in Section~\ref{sec:concl}.

\section{Misaligned Be star disk simulation}
\label{sec:bedisk}

We use the smoothed particle hydrodynamics (SPH) code {\sc phantom} \citep{Price2010,Lodato2010} to model the system composed of a Be star surrounded by a highly misaligned accretion disk with initial inclination $i=70\dg$ and a companion neutron star. 
The binary components are modelled as sink particles with masses of $M_*=18\,M_{\odot}$ and $M_{\rm ns}=1.4\,M_{\odot}$, respectively. 
We choose a binary separation of $a=95\,R_{\odot}$ that corresponds to an orbital period of $P_{\rm b}=24$ days and an eccentricity of $e=0.34$. 

These values correspond to the orbital parameters of the most active Be/X-ray binary transient 4U 0115+63. This is one of the first discovered \citep{Giacconi1972,Whitlock1989} and extensively observed \citep{Campana1996,Negueruela1997} Be/X-ray binaries. Thanks to its very well constrained orbital parameters, many models for Be/X-ray binaries are based on this source \citep{Negueruela2001,Okazaki2001}. Giant outbursts in this system have been associated with the presence of a tilted disk around the Be star \citep{Reig2007b,martin2011,Kato2014} and the timescale between subsequent giant outbursts has been estimated to be 3 yrs, even though sometimes these outbursts can occur 1-1.5 yrs apart \citep{Reig2007b,Reig2018}.
Note that, for this system, the ZLK oscillations of the Be star disk might operate on a timescale that matches the observed $3\,\rm yr$ frequency of type II outbursts, provided that
the disk is not completely destroyed during the outburst \citep{martinfranchini2019}. 
 
The accretion radii are chosen to be $R_{\rm acc,*}=8\,R_{\odot}$ and $R_{\rm acc,ns}=1\,R_{\odot}$  for the Be star and the neutron star respectively. Particles inside these radii are accreted onto the respective sink particle and their mass and angular momentum added to the sink \citep{bate1995}.
The accretion disk around the Be star extends initially from $R_{\rm in}=8\,R_{\odot}$ to an outer radius $R_{\rm out}=50\,R_{\odot}$ and has a total initial mass $M_{\rm d}=1\times10^{-8}M_{\odot}$, distributed with a surface density profile $\Sigma \propto R^{-1}$ \citep{Cote1987,Porter1999}. The neutron star does not have an accretion disk at the beginning of the simulation.

The Be star accretion disk initially expands out to the tidal truncation radius. Note that given the moderate binary eccentricities of these systems, the gravitational effect of the neutron star depends on the orbital phase. Therefore the disk radius and the truncation radius are also phase dependent. 
As expected, the Be star disk is smaller when the neutron star is at periastron and the truncation radius reaches its minimum value, while it spreads outwards when the neutron star is at apastron \citep{Goldreich1979,Artymowicz1994,Okazaki2001,Negueruela2001}.

We use $N=1 \times 10^6$ particles to resolve the Be star gaseous disk in our simulation. 
We model the disk viscosity using a Shakura-Sunyaev $\alpha$ viscosity which is implemented via a direct Navier-Stokes viscosity \citep{flebbe1994}. We therefore use the switches for shocks and to prevent particle interpenetration \citep{cullendehnen2010} with $\alpha_{\rm AV}$ between 0.1 and 1 and $\beta_{\rm AV}=2$ to minimize the amount of numerical viscosity in the simulation \citep{meru2012}.
We choose $\alpha=0.2$ to model the Be star accretion disk \citep{King2007,rimulo2018,martinetal2019}.

We use the locally isothermal equation of state for the gas adopted in \citep{Farris2014} 
\begin{equation}
c_{\rm s} =  \mathcal{F} \,c_{\rm s0} \left( \frac{a}{M_* +M_{\rm ns}}\right)^{q} \left(\frac{M_*}{R_*} + \frac{M_{\rm ns}}{R_{\rm ns}}\right)^{q},
\label{eq:farris}
\end{equation}
where $R_*,\,R_{\rm ns}$ are the radial distances of the gas particles from the Be star and the neutron star respectively and $c_{\rm s0}$ is a constant of proportionality. We reduce the temperature of the gas around the neutron star through the constant $\mathcal{F}$ by two orders of magnitude at radial distance less than $35\,R_{\odot}$ from the neutron star, similar to \cite{Smallwood2021}. This is necessary to better resolve the disk around the neutron star since the viscosity in SPH depends on the number of particles and is therefore prohibitively large for the particles to form a proper disk without being promptly accreted onto the neutron star. Reducing the temperature around the neutron star reduces the disk aspect ratio and therefore the viscosity.
We take $q=0$ to model a flared circumstellar disk.
\cite{martinfranchini2019} showed that flared circumstellar disks are unstable against ZLK oscillations for binaries with orbital period up to roughly 100 days.

According to the analytical calculations outlined in \cite{martinfranchini2019}, since the viscous timescale must be short enough for the disk to spread outwards between outbursts the aspect ratio at the disk outer edge should be $\geq 0.06$. At the same time the aspect ratio at the disk outer edge cannot be too large ($H/R \leq 0.11$), otherwise the disk would be stable against ZLK oscillations \citep{lo2017}.
Therefore we take the disk aspect ratio at the inner edge to be $H/R\,(R_{\rm in}) = 0.024$ which corresponds to $H/R\,(R=50\,\rm R_\odot)=0.06 $.

\begin{figure}
\includegraphics[width=\columnwidth]{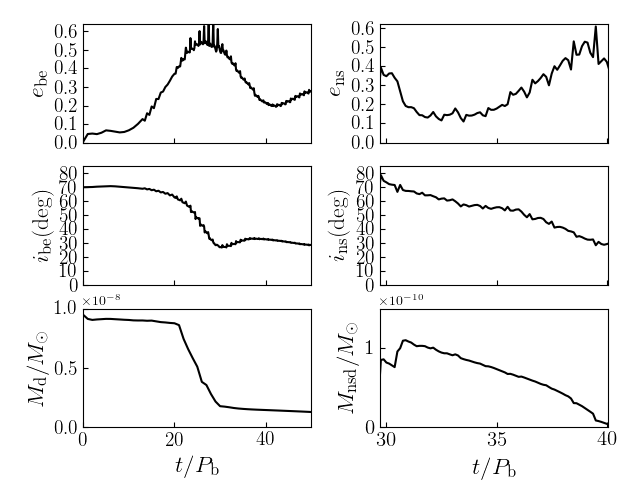}
\caption{Be star disk simulation. Eccentricity (upper panels), inclination (middle panels) and disk mass (lower panels) evolution of the Be star disk (left) and neutron star disk (right) averaged over the whole disk. The time is in units of the binary period $P_{\rm b}$. The initial inclination of the Be star disk is $i=70^{\circ}$.}
\label{fig:params_be}
\end{figure}

\begin{figure}
    \includegraphics[width=\columnwidth]{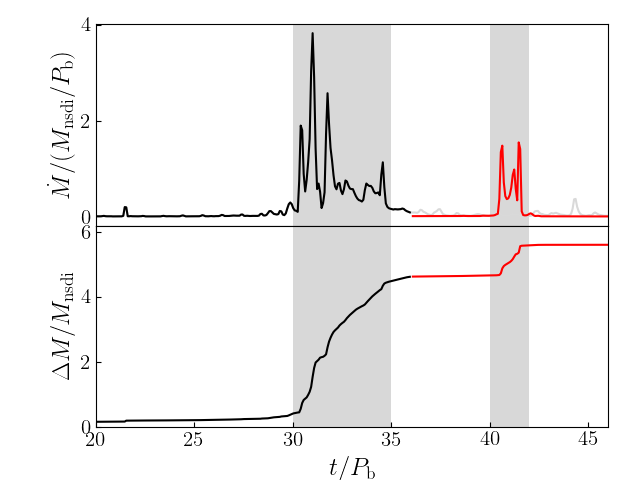} 
\caption{Upper panel: accretion rate onto the neutron star (at an accretion radius of $1\,R_{\odot}$) in units of the newly formed neutron star disk mass, measured at $t=36\,P_{\rm b}$, and of the binary orbital period. Bottom panel: total accreted mass onto the neutron star in units of $M_{\rm nsdi}$. The black lines show the results of the Be star disk simulation (Section \ref{sec:bedisk}) while the red lines show the results of the simulation with the neutron star disk only (Section~\ref{sec:nsdisk}).  }
\label{fig:mdot_be}
\end{figure}

\begin{figure}
    \includegraphics[width=\columnwidth]{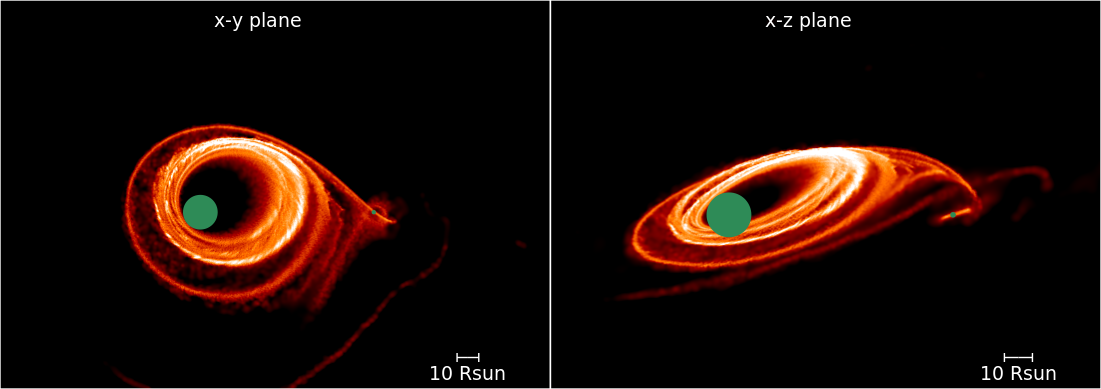}
    \includegraphics[width=\columnwidth]{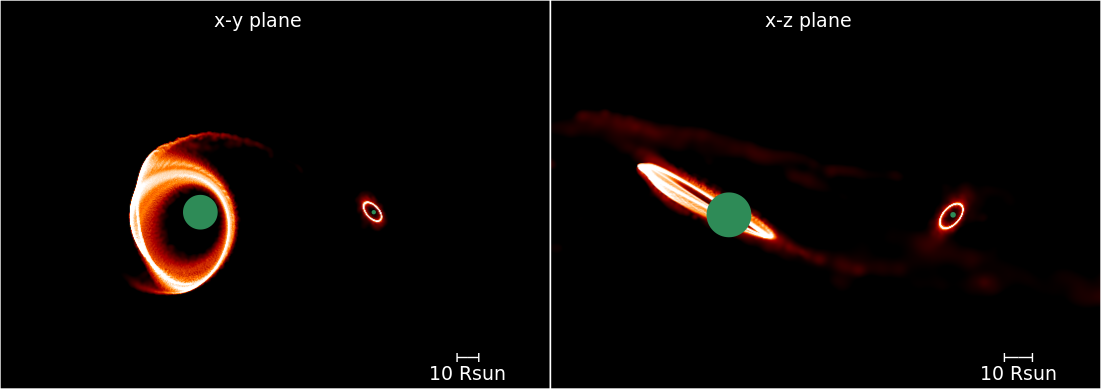}
    \includegraphics[width=\columnwidth]{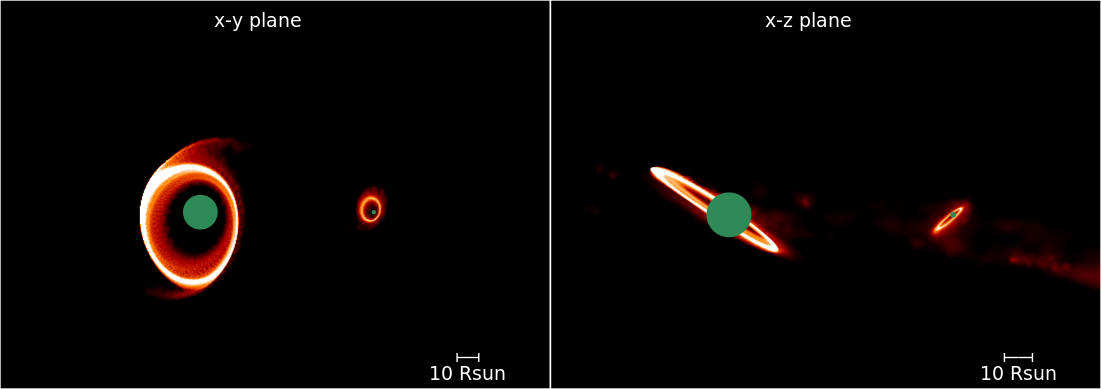}
    \caption{Column density plots of the Be star and neutron star disks. The binary components are shown in green and their size is determined by their accretion radii.  The view is of the $x$-$y$ plane (i.e. the binary orbital plane) (left panels) and of the $x-z$ plane (right panels). The color scale spans about three orders of magnitude in density and is the same for all the plots. The first, second and third row correspond to $29,\,33,\,38\,P_{\rm b}$. }
    \label{fig:snap}
\end{figure}

Figure \ref{fig:params_be} shows the eccentricity and inclination evolution of the Be star accretion disk (left) and of the material that forms a disk around the neutron star (right) averaged over the radial extent of each respective disk.
At a time of roughly $30\,P_{\rm b}$ the Be star disk undergoes a large ZLK oscillation and its eccentricity increases up to about $e=0.6$. The disk fills the Roche lobe and mass is transferred to the neutron star. This is reflected by the Be star disk mass evolution, shown in the bottom left panel of Fig. \ref{fig:params_be}. 
Since the Be star disk is misaligned with respect to the binary plane,  the mass transfer to the neutron star does not necessarily occur in the binary orbital plane, but instead it can take place on a highly misaligned orbit (see upper panels in Fig. \ref{fig:snap}). 
An accretion disk with mass $\sim 10^{-10}\,M_{\odot}$ forms around the neutron star at a time of around $30\,P_{\rm b}$ with initial eccentricity $e\simeq 0.4$ and inclination $\sim 70^{\circ}$. While the disk is poorly resolved, its evolution is shown in the right panels of Fig.~\ref{fig:params_be}. The neutron star disk initially circularises to an eccentricity of about 0.1 over the first few orbital periods. Because the neutron star disk forms above the critical angle required for ZLK oscillations, it is also unstable  and its eccentricity increases  up to a maximum of about $e=0.6$ at a time of around $38\,P_{\rm b}$.
However we can only resolve this newly formed disk for about $12\,P_{\rm b}$ since the disk is accreted rapidly during the ZLK eccentricity peak. 

The black lines in the left panels of Figure \ref{fig:mdot_be} show the accretion rate (upper panel) and the total accreted mass (lower panel) onto the neutron star as a function of time in units of the binary orbital period.
Both quantities are scaled to the newly formed neutron star disk mass measured at $t=36\,P_{\rm b}$, i.e. $M_{\rm nsdi}=8\times10^{-11}M_{\odot}$.
At $30\,P_{\rm b}$ the Be star disk undergoes the first large ZLK oscillation causing a high amplitude, longer accretion episode that resembles a Type II outburst. This is due to the mass transfer between the Be star disk and the neutron star, which occurs via Roche lobe overflow \citep{martin2014be,franchini2019a}.

Figure \ref{fig:snap} shows the column density maps of the simulation at times of $29,\,33,\,38\,P_{\rm b}$ from the upper to the lower panel in the $x$-$y$ (left panels) and $x-z$ plane (right panels). We can see the neutron star disk formation as a consequence of the first ZLK oscillation of the Be star disk in the upper panel. The neutron star disk becomes almost circular in the middle panel (consistent with the upper right panel of Fig. \ref{fig:params_be}). The neutron star disk  eccentricity increases again due to the ZLK mechanism at around $38\,P_{\rm b}$ (lower panel).

\section{Eccentric neutron star disk simulation}
\label{sec:nsdisk}

Previous hydrodynamical simulations performed by \cite{Hayasaki2004,Hayasaki2006} investigated the accretion flow around the neutron star induced by both Roche lobe overflow and from the pericentre passage. They found that a time persistent accretion disk forms around the compact object in both cases.
However in their simulations the mass flow onto the neutron star is assumed to be coplanar with the binary orbital plane.

Even though we tuned the resolution around the neutron star in the simulation by changing the equation of state, this was not sufficient to resolve accurately the neutron star disk. 
Thus, in order to understand the neutron star disk dynamics, we now consider a second simulation in which we take the same binary parameters but have no disk around the Be star and an initially circular accretion disk inclined by $70^{\circ}$ around the neutron star modelled with $N=500,000$ particles. The initial disk mass is $M_{\rm d}=1\times10^{-8}M_{\odot}$ distributed with a surface density profile $\Sigma \propto R^{-3/2}$ between $R_{\rm in}=3\,R_{\odot}$ and $R_{\rm out}=10\,R_{\odot}$. 
Note that, since the disk mass is very small compared to the mass of the binary the choice of its initial value does not influence the dynamics of the neutron star disk. In addition, we scale the accretion rate and the total accreted mass to the initial neutron star disk mass. 
The accretion radii of the two sinks are the same as in the previous simulation. The disk aspect ratio is $H/R\,(R_{\rm in}) = 0.01$ and the viscosity parameter is $\alpha=0.1$. We modelled the neutron star disk using a global isothermal equation of state, i.e. we use Eq. \ref{eq:farris} with $q=0$ and $\mathcal{F}=1$ in the entire domain. We chose the initial inclination and the disk radial extent based on the characteristics of the neutron star disk that forms from the Be star disk ZLK oscillations in the previous simulation (see right panel of Fig. \ref{fig:params_be}). 
The timescale for ZLK oscillations depends on the disk radial extent and on the distribution of mass within it \citep{martin2014be,martinfranchini2019}. In particular, larger disk outer radii and higher mass contents in the outer regions lead to shorter timescales.

The red lines in Figure \ref{fig:mdot_be} show the accretion rate (upper panel) and the accreted mass (lower panel) onto the neutron star in this second simulation. 
These quantities are both scaled to the initial mass of the neutron star disk in this second simulation, i.e. $10^{-8}M_{\odot}$. 
We start the red lines at $t=36\,P_{\rm b}$ since by this time the first ZLK oscillation of the Be star disk ceases to cause mass transfer to the neutron star disk. The dynamics of the neutron star disk are dominated by the inflow of material until this time. Furthermore the material in the disk at this time is on  almost circular orbits (see upper right panel in Fig. \ref{fig:params_be}), which corresponds to the initial condition in our second simulation.

The neutron star disk is initially circular but rapidly, within the first few binary orbits, reaches an eccentricity of $e=0.8$ while its inclination drops to $\sim 39\dg$ due to the ZLK mechanism. We find this eccentricity increase to lead to an episode of enhanced accretion rate onto the neutron star, represented by the peak at $40\,P_{\rm b}$, shown by the red line in the upper right panel of Fig. \ref{fig:mdot_be}. 
The characteristics of this second peak are compatible with type II outbursts since the accretion rate increases by more than an order of magnitude and the peak lasts for many orbital periods.
This second outburst amplitude is about a factor of more than 2 smaller than the first outburst seen in the Be star disk simulation.     
The relative magnitude of the two outbursts essentially depends on how much mass is directly accreted with compared to the mass that is transferred and forms the disk. However, since we take the initial neutron star disk mass from the poorly resolved material in the Be star disk simulation, what we show is likely a lower limit. If we were able to resolve the neutron star disk better in the first simulation we would expect a higher disk mass and therefore a larger second outburst.

The timescale for the neutron star disk ZLK oscillation depends on the disk viscosity and on the distribution of the gas within the disk. If the disk is viscous enough to expand rapidly, the type II outburst occurs after a few binary orbits because the disk outer parts become ZLK unstable fast.
The timescale for global ZLK oscillations can be written as \citep{martin2014be,martinfranchini2019}
\begin{equation}
    \frac{\langle \tau_{\rm ZLK}\rangle}{P_{\rm b}} \approx \frac{\sqrt{M\,M_1}}{M_2}\left(\frac{a}{R_{\rm in}}\right)^{3/2} \frac{4-p}{5/2-p}\frac{(R_{\rm out}/R_{\rm in})^{2.5-p}-1}{(R_{\rm out}/R_{\rm in})^{4-p}-1}
    \label{eq:ZLKtime}
\end{equation}
where $p$ is the surface density power law index, $M$ is the total mass of the binary and $M_1,M_2$ are the Be star or neutron star disc masses depending on which disc is undergoing ZLK oscillations. 

The neutron star disk in the previous simulation (see Section \ref{sec:bedisk}) forms with an outer radius of approximately $13\,R_{\odot}$. The corresponding global ZLK timescale is, assuming a surface density profile $\Sigma \propto R^{-3/2}$, $\langle \tau_{\rm ZLK}\rangle \simeq 11\,P_{\rm b}$ \citep{martin2014be,martinfranchini2019}. 
We note that the surface density profile of the neutron star disc might deviate from the initial power law as a result of the interaction with the companion and of its own viscous evolution. A steeper (shallower) profile leads to a longer (shorter) ZLK timescale (see Eq. \ref{eq:ZLKtime}).
The eccentricity oscillation shown in the upper right panel of Figure \ref{fig:params_be} occurs at around $\sim 3\,P_{\rm b}$ which is slightly shorter than the predicted $\langle \tau_{\rm ZLK}\rangle/2$. Since the disk is initially eccentric, the ZLK period is indeed expected to be shorter than the analytical prediction \citep{franchini2019a}.

The neutron star disk in this second simulation reaches instead the peak eccentricity at $6\,P_{\rm b}$ which is consistent with the theoretical estimate of the ZLK timescale for an initially circular disk based on its outer radius and surface density profiles, i.e. $\langle \tau_{\rm ZLK}\rangle \simeq 11\,P_{\rm b}$.

The recurrence time between subsequent type II outbursts has been estimated to be around 3 yrs from observations \citep{Whitlock1989}. This periodicity has also been confirmed numerically by  \cite{Negueruela2001b}. However \cite{Reig2007b} showed that this three-years cycle might be interrupted by two close, 1-1.5 yrs apart, type II outbursts  When this occurs, after the second outburst the disk might be completely dissipated \citep{Reig2018}. 
For the system 4U 0115+63 this translates to $\sim 15-23\,P_{\rm b}$ between the two outbursts.

The timescale between the two outbursts corresponds to the time to form the neutron star disk (about half the ZLK timescale of the Be star disk) plus half the ZLK oscillation timescale of the neutron star disk in the second simulation. The analytically estimated value is consistent with the timescale we show in the upper panel of Figure \ref{fig:mdot_be}, i.e. $\sim 12\,P_{\rm b}$.
This is slightly shorter than the expected time for 4U 0115+63. However, since the ZLK period is very sensitive to the disk surface density profile, a slightly steeper profile, e.g. $\Sigma \propto R^{-2}$, leads to a longer ZLK timescale.  The estimated timescale for the ZLK oscillation of the neutron star disk for such profile is $\langle \tau_{\rm ZLK}\rangle/2\simeq 7\,P_{\rm b}$. This added to the ZLK timescale of the Be star disk, results in a longer interval between outbursts, closer to the 1 yr interval estimated for this source. 

Furthermore, the circularization radius of the material coming from the Be star disk, i.e. the neutron star disk outer radius, might also depend on the viscosity, equation of state and mass distribution in the simulation. In particular, smaller outer radii lead to longer ZLK oscillation timescales. 
Therefore in order for the type II outbursts to be further apart we expect a steeper surface density profile or a much smaller outer radius of the neutron star disk.
The timescale for the first ZLK oscillation peak is also longer with a smaller initial disk inclination or a smaller disk aspect ratio \citep{fu2015a}. Thus, while our simulation shows a timescale that is a little short for the specific case of 4U 0115+63, we suggest that the neutron star disk might form with a lower inclination than $70^\circ$.

Multiple type II outbursts have been also observed in the X-ray and optical light curves of the two very well studied sources 1A 0535+262 \citep{Rosenberg1975} and SAX J2103.5+4545 \citep{Hulleman1998}.

1A 0535+262 has an inferred orbital period $P_{\rm b}=111$ days \citep{Finger1994} and underwent the first type II outburst after 10 yrs of quiescence in 2005 \citep{smith2005}, followed by another one in 2009 \citep{Caballero2010a} and a third one 2 yrs later in 2011 \citep{Monageng2017}. The interval between the 2005 and 2009 outbursts corresponds to $\sim 13\,P_{\rm b}$ for this particular source and is consistent with the timescale inferred from our simulations while the 2011 outburst occured after $\sim 7\,P_{\rm b}$.

The source SAX J2103.5+4545 has one of the shortest known Be/X-ray binaries orbital period at $P_{\rm b}=12.68$ days \citep{Baykal2000}. In the optical band, this source showed two type II outbursts 2.7 yrs apart between 2007 and 2010 and then another pair 0.7 yrs apart \citep{Camero2014}. As these timescales translate to $\sim 75$ and $\sim 19\,P_{\rm b}$, our model could in principle only explain the latest pair of outbursts. 

\section{Conclusions}
\label{sec:concl}

In this paper we have investigated the dynamics of a highly misaligned disk surrounding the neutron star in a Be/X-ray binary system using 3D SPH simulations.
We first ran a simulation of a Be star disk and showed that during ZLK oscillations of eccentricity and inclination, mass transfer produces a highly inclined eccentric neutron star accretion disk.
We were unable to obtain a good enough resolution of the neutron star disk by forming it in this way and so we therefore used the parameters of this newly formed neutron star disk as the initial conditions for a second simulation involving only the neutron star disk.

We have shown that the dynamics of the two disks can lead to type-II outburst pairs. The first type II outburst is triggered by the transfer of Be star disk material to the neutron star. The subsequent excitation of the neutron star disk eccentricity, as a result of the ZLK mechanism, causes another large increase of accretion rate onto the compact object. We interpret this as a possible second type II outburst. We finally find the timescale between the two giant outbursts to be $\sim 12\,P_{\rm b}$, compatible with the $1-1.5$ yrs interval observed in 4U~0115+63 \citep{Reig2007b,Reig2018} and possibly also with outbursts pairs in the sources 1A 0535+262 and SAX J2103.5+4545.
A more thorough exploration of the parameter space for this mechanism to occur in other Be/X-ray binaries is deferred to a future work.

It is worth noting that whether the Be star disc driven type II outburst occurs as a single event or produces a pair of outbursts does not influence the occurrence rate of type II outbursts in general, as the pair is driven by the same mass transfer mechanism and the neutron star disc driven outburst does not influence the Be star accretion disc. We can therefore say that the occurrence rates of type II outbursts (or outburst pairs) driven by ZLK oscillations are the rates estimated previously in \cite{martinfranchini2019} and depend on the discs viscosity and temperature, that essentially regulate the mass distribution within the disc, as well as the timescale for Be star disc refilling \citep[e.g.][]{Suffak2022}.

We finally note that the resolution in the first simulation was not high enough to completely resolve the second outburst due to the eccentric neutron star disk. This might due to the very large viscosity caused by the small number of particles in the disk coupled with the fact that type I outbursts continue to occur because of the presence of the Be star disk. The improvement of the resolution of the disk around the neutron star using more advanced numerical techniques will be the subject of a future work.

\begin{acknowledgments}
We warmly thank the referee for very useful comments that improved the quality of the manuscript.
AF acknowledges financial support provided under the European Union's H2020 ERC Consolidator Grant ``Binary Massive Black Hole Astrophysics" (B Massive, Grant Agreement: 818691).
RGM acknowledges support from NASA through grant 80NSSC21K0395.
We acknowledge the use of SPLASH \citep{Price2007} for the rendering of the figures.

\end{acknowledgments}

\section*{Data Availability}

The SPH simulations results in this paper can be reproduced using the {\sc phantom} code (Astrophysics Source Code Library identifier {\tt ascl.net/1709.002}).  The data underlying this article will be shared on reasonable request to the corresponding author.

\bibliography{references_sorted}{}

\newcommand{\noop}[1]{}
\begin{thebibliography}{}
\expandafter\ifx\csname natexlab\endcsname\relax\def\natexlab#1{#1}\fi
\providecommand{\url}[1]{\href{#1}{#1}}
\providecommand{\dodoi}[1]{doi:~\href{http://doi.org/#1}{\nolinkurl{#1}}}
\providecommand{\doeprint}[1]{\href{http://ascl.net/#1}{\nolinkurl{http://ascl.net/#1}}}
\providecommand{\doarXiv}[1]{\href{https://arxiv.org/abs/#1}{\nolinkurl{https://arxiv.org/abs/#1}}}

\bibitem[{{Artymowicz} \& {Lubow}(1994)}]{Artymowicz1994}
{Artymowicz}, P., \& {Lubow}, S.~H. 1994, \apj, 421, 651,
  \dodoi{10.1086/173679}

\bibitem[{{Bate} {et~al.}(1995){Bate}, {Bonnell}, \& {Price}}]{bate1995}
{Bate}, M.~R., {Bonnell}, I.~A., \& {Price}, N.~M. 1995, \mnras, 277, 362,
  \dodoi{10.1093/mnras/277.2.362}

\bibitem[{{Baykal} {et~al.}(2000){Baykal}, {Stark}, \& {Swank}}]{Baykal2000}
{Baykal}, A., {Stark}, M.~J., \& {Swank}, J. 2000, \apjl, 544, L129,
  \dodoi{10.1086/317320}

\bibitem[{{Caballero} {et~al.}(2010){Caballero}, {Kretschmar}, {Pottschmidt},
  {Santangelo}, {Wilms}, {Kreykenbohm}, {Ferrigno}, {Suchy}, {Rothschild},
  {Finger}, {Postnov}, {McBride}, {Domingo}, {Sch{\"o}nherr}, {Klochkov},
  {Staubert}, \& {Camero-Arranz}}]{Caballero2010a}
{Caballero}, I., {Kretschmar}, P., {Pottschmidt}, K., {et~al.} 2010, in
  American Institute of Physics Conference Series, Vol. 1248, X-ray Astronomy
  2009; Present Status, Multi-Wavelength Approach and Future Perspectives, ed.
  A.~{Comastri}, L.~{Angelini}, \& M.~{Cappi}, 147--148,
  \dodoi{10.1063/1.3475172}

\bibitem[{{Camero} {et~al.}(2014){Camero}, {Zurita}, {Guti{\'e}rrez-Soto},
  {{\"O}zbey Arabac{\i}}, {Nespoli}, {Kiaeerad}, {Beklen}, {Garc{\'\i}a-Rojas},
  \& {Caballero-Garc{\'\i}a}}]{Camero2014}
{Camero}, A., {Zurita}, C., {Guti{\'e}rrez-Soto}, J., {et~al.} 2014, \aap, 568,
  A115, \dodoi{10.1051/0004-6361/201423452}

\bibitem[{{Campana}(1996)}]{Campana1996}
{Campana}, S. 1996, \apss, 239, 113, \dodoi{10.1007/BF00653771}

\bibitem[{{Cote} \& {Waters}(1987)}]{Cote1987}
{Cote}, J., \& {Waters}, L.~B.~F.~M. 1987, \aap, 176, 93

\bibitem[{{Cullen} \& {Dehnen}(2010)}]{cullendehnen2010}
{Cullen}, L., \& {Dehnen}, W. 2010, \mnras, 408, 669,
  \dodoi{10.1111/j.1365-2966.2010.17158.x}

\bibitem[{{Farris} {et~al.}(2014){Farris}, {Duffell}, {MacFadyen}, \&
  {Haiman}}]{Farris2014}
{Farris}, B.~D., {Duffell}, P., {MacFadyen}, A.~I., \& {Haiman}, Z. 2014, \apj,
  783, 134, \dodoi{10.1088/0004-637X/783/2/134}

\bibitem[{{Finger} {et~al.}(1994){Finger}, {Cominsky}, {Wilson}, {Harmon}, \&
  {Fishman}}]{Finger1994}
{Finger}, M.~H., {Cominsky}, L.~R., {Wilson}, R.~B., {Harmon}, B.~A., \&
  {Fishman}, G.~J. 1994, in American Institute of Physics Conference Series,
  Vol. 308, The Evolution of X-ray Binariese, ed. S.~{Holt} \& C.~S. {Day},
  459, \dodoi{10.1063/1.46032}

\bibitem[{{Finger} {et~al.}(1996){Finger}, {Wilson}, \& {Harmon}}]{Finger1996}
{Finger}, M.~H., {Wilson}, R.~B., \& {Harmon}, B.~A. 1996, \apj, 459, 288,
  \dodoi{10.1086/176892}

\bibitem[{{Flebbe} {et~al.}(1994){Flebbe}, {Muenzel}, {Herold}, {Riffert}, \&
  {Ruder}}]{flebbe1994}
{Flebbe}, O., {Muenzel}, S., {Herold}, H., {Riffert}, H., \& {Ruder}, H. 1994,
  \apj, 431, 754, \dodoi{10.1086/174526}

\bibitem[{{Franchini} \& {Martin}(2019)}]{franchini2019b}
{Franchini}, A., \& {Martin}, R.~G. 2019, \apjl, 881, L32,
  \dodoi{10.3847/2041-8213/ab3920}

\bibitem[{{Franchini} {et~al.}(2019){Franchini}, {Martin}, \&
  {Lubow}}]{franchini2019a}
{Franchini}, A., {Martin}, R.~G., \& {Lubow}, S.~H. 2019, \mnras, 485, 315,
  \dodoi{10.1093/mnras/stz424}

\bibitem[{{Fu} {et~al.}(2015{\natexlab{a}}){Fu}, {Lubow}, \&
  {Martin}}]{fu2015b}
{Fu}, W., {Lubow}, S.~H., \& {Martin}, R.~G. 2015{\natexlab{a}}, \apj, 813,
  105, \dodoi{10.1088/0004-637X/813/2/105}

\bibitem[{{Fu} {et~al.}(2015{\natexlab{b}}){Fu}, {Lubow}, \&
  {Martin}}]{fu2015a}
---. 2015{\natexlab{b}}, \apj, 807, 75, \dodoi{10.1088/0004-637X/807/1/75}

\bibitem[{{Giacconi} {et~al.}(1972){Giacconi}, {Murray}, {Gursky}, {Kellogg},
  {Schreier}, \& {Tananbaum}}]{Giacconi1972}
{Giacconi}, R., {Murray}, S., {Gursky}, H., {et~al.} 1972, \apj, 178, 281,
  \dodoi{10.1086/151790}

\bibitem[{{Goldreich} \& {Tremaine}(1979)}]{Goldreich1979}
{Goldreich}, P., \& {Tremaine}, S. 1979, \apj, 233, 857, \dodoi{10.1086/157448}

\bibitem[{{Hayasaki} \& {Okazaki}(2004)}]{Hayasaki2004}
{Hayasaki}, K., \& {Okazaki}, A.~T. 2004, \mnras, 350, 971,
  \dodoi{10.1111/j.1365-2966.2004.07702.x}

\bibitem[{{Hayasaki} \& {Okazaki}(2006)}]{Hayasaki2006}
---. 2006, \mnras, 372, 1140, \dodoi{10.1111/j.1365-2966.2006.10917.x}

\bibitem[{{Hulleman} {et~al.}(1998){Hulleman}, {in 't Zand}, \&
  {Heise}}]{Hulleman1998}
{Hulleman}, F., {in 't Zand}, J.~J.~M., \& {Heise}, J. 1998, \aap, 337, L25.
\newblock \doarXiv{astro-ph/9807280}

\bibitem[{{Kato}(2014)}]{Kato2014}
{Kato}, S. 2014, \pasj, 66, 74, \dodoi{10.1093/pasj/psu050}

\bibitem[{{King} {et~al.}(2007){King}, {Pringle}, \& {Livio}}]{King2007}
{King}, A.~R., {Pringle}, J.~E., \& {Livio}, M. 2007, \mnras, 376, 1740,
  \dodoi{10.1111/j.1365-2966.2007.11556.x}

\bibitem[{{Kozai}(1962)}]{Kozai1962}
{Kozai}, Y. 1962, \aj, 67, 591, \dodoi{10.1086/108790}

\bibitem[{{Kretschmar} {et~al.}(2012){Kretschmar}, {Nespoli}, {Reig}, \&
  {Anders}}]{Kretschmar2012}
{Kretschmar}, P., {Nespoli}, E., {Reig}, P., \& {Anders}, F. 2012, in
  Proceedings of ``An INTEGRAL view of the high-energy sky (the first 10
  years)'', 16.
\newblock \doarXiv{1302.3434}

\bibitem[{{Lidov}(1962)}]{Lidov1962}
{Lidov}, M.~L. 1962, \planss, 9, 719, \dodoi{10.1016/0032-0633(62)90129-0}

\bibitem[{{Lodato} \& {Price}(2010)}]{Lodato2010}
{Lodato}, G., \& {Price}, D.~J. 2010, \mnras, 405, 1212,
  \dodoi{10.1111/j.1365-2966.2010.16526.x}

\bibitem[{{Lubow} \& {Ogilvie}(2017)}]{lo2017}
{Lubow}, S.~H., \& {Ogilvie}, G.~I. 2017, \mnras, 469, 4292,
  \dodoi{10.1093/mnras/stx990}

\bibitem[{{Martin} \& {Franchini}(2019)}]{martinfranchini2019}
{Martin}, R.~G., \& {Franchini}, A. 2019, \mnras, 489, 1797,
  \dodoi{10.1093/mnras/stz2250}

\bibitem[{{Martin} \& {Franchini}(2021)}]{martinfranchini2021}
---. 2021, arXiv e-prints, arXiv:2111.08642.
\newblock \doarXiv{2111.08642}

\bibitem[{{Martin} {et~al.}(2014{\natexlab{a}}){Martin}, {Nixon}, {Armitage},
  {Lubow}, \& {Price}}]{martin2014be}
{Martin}, R.~G., {Nixon}, C., {Armitage}, P.~J., {Lubow}, S.~H., \& {Price},
  D.~J. 2014{\natexlab{a}}, \apj, 790, L34, \dodoi{10.1088/2041-8205/790/2/L34}

\bibitem[{{Martin} {et~al.}(2014{\natexlab{b}}){Martin}, {Nixon}, {Lubow},
  {Armitage}, {Price}, {Do{\u{g}}an}, \& {King}}]{martin2014}
{Martin}, R.~G., {Nixon}, C., {Lubow}, S.~H., {et~al.} 2014{\natexlab{b}},
  \apjl, 792, L33, \dodoi{10.1088/2041-8205/792/2/L33}

\bibitem[{{Martin} {et~al.}(2019){Martin}, {Nixon}, {Pringle}, \&
  {Livio}}]{martinetal2019}
{Martin}, R.~G., {Nixon}, C.~J., {Pringle}, J.~E., \& {Livio}, M. 2019, \na,
  70, 7, \dodoi{10.1016/j.newast.2019.01.001}

\bibitem[{{Martin} {et~al.}(2011){Martin}, {Pringle}, {Tout}, \&
  {Lubow}}]{martin2011}
{Martin}, R.~G., {Pringle}, J.~E., {Tout}, C.~A., \& {Lubow}, S.~H. 2011,
  \mnras, 416, 2827, \dodoi{10.1111/j.1365-2966.2011.19231.x}

\bibitem[{{Meru} \& {Bate}(2012)}]{meru2012}
{Meru}, F., \& {Bate}, M.~R. 2012, \mnras, 427, 2022,
  \dodoi{10.1111/j.1365-2966.2012.22035.x}

\bibitem[{{Monageng} {et~al.}(2017){Monageng}, {McBride}, {Coe}, {Steele}, \&
  {Reig}}]{Monageng2017}
{Monageng}, I.~M., {McBride}, V.~A., {Coe}, M.~J., {Steele}, I.~A., \& {Reig},
  P. 2017, \mnras, 464, 572, \dodoi{10.1093/mnras/stw2354}

\bibitem[{{Negueruela}(1998)}]{Negueruela1998}
{Negueruela}, I. 1998, \aap, 338, 505.
\newblock \doarXiv{astro-ph/9807158}

\bibitem[{{Negueruela} \& {Okazaki}(2001)}]{Negueruela2001}
{Negueruela}, I., \& {Okazaki}, A.~T. 2001, \aap, 369, 108,
  \dodoi{10.1051/0004-6361:20010146}

\bibitem[{{Negueruela} {et~al.}(2001){Negueruela}, {Okazaki}, {Fabregat},
  {Coe}, {Munari}, \& {Tomov}}]{Negueruela2001b}
{Negueruela}, I., {Okazaki}, A.~T., {Fabregat}, J., {et~al.} 2001, \aap, 369,
  117, \dodoi{10.1051/0004-6361:20010077}

\bibitem[{{Negueruela} {et~al.}(1997){Negueruela}, {Grove}, {Coe}, {Fabregat},
  {Finger}, {Phlips}, {Roche}, {Steele}, \& {Unger}}]{Negueruela1997}
{Negueruela}, I., {Grove}, J.~E., {Coe}, M.~J., {et~al.} 1997, \mnras, 284,
  859, \dodoi{10.1093/mnras/284.4.859}

\bibitem[{{Okazaki} {et~al.}(2013){Okazaki}, {Hayasaki}, \&
  {Moritani}}]{Okazaki2013}
{Okazaki}, A.~T., {Hayasaki}, K., \& {Moritani}, Y. 2013, \pasj, 65, 41,
  \dodoi{10.1093/pasj/65.2.41}

\bibitem[{{Okazaki} \& {Negueruela}(2001)}]{Okazaki2001}
{Okazaki}, A.~T., \& {Negueruela}, I. 2001, \aap, 377, 161,
  \dodoi{10.1051/0004-6361:20011083}

\bibitem[{{Porter}(1999)}]{Porter1999}
{Porter}, J.~M. 1999, \aap, 348, 512.
\newblock \doarXiv{astro-ph/9906381}

\bibitem[{{Porter} \& {Rivinius}(2003)}]{Porter2003}
{Porter}, J.~M., \& {Rivinius}, T. 2003, \pasp, 115, 1153,
  \dodoi{10.1086/378307}

\bibitem[{{Price}(2007)}]{Price2007}
{Price}, D.~J. 2007, \pasa, 24, 159, \dodoi{10.1071/AS07022}

\bibitem[{{Price} \& {Federrath}(2010)}]{Price2010}
{Price}, D.~J., \& {Federrath}, C. 2010, \mnras, 406, 1659,
  \dodoi{10.1111/j.1365-2966.2010.16810.x}

\bibitem[{{Pringle}(1991)}]{Pringle1991}
{Pringle}, J.~E. 1991, MNRAS, 248, 754

\bibitem[{{Rajoelimanana} {et~al.}(2011){Rajoelimanana}, {Charles}, \&
  {Udalski}}]{Rajoelimanana2011}
{Rajoelimanana}, A.~F., {Charles}, P.~A., \& {Udalski}, A. 2011, \mnras, 413,
  1600, \dodoi{10.1111/j.1365-2966.2011.18243.x}

\bibitem[{{Reig} \& {Blinov}(2018)}]{Reig2018}
{Reig}, P., \& {Blinov}, D. 2018, \aap, 619, A19,
  \dodoi{10.1051/0004-6361/201833649}

\bibitem[{{Reig} {et~al.}(2007){Reig}, {Larionov}, {Negueruela}, {Arkharov}, \&
  {Kudryavtseva}}]{Reig2007b}
{Reig}, P., {Larionov}, V., {Negueruela}, I., {Arkharov}, A.~A., \&
  {Kudryavtseva}, N.~A. 2007, \aap, 462, 1081,
  \dodoi{10.1051/0004-6361:20066217}

\bibitem[{{R{\'{\i}}mulo} {et~al.}(2018){R{\'{\i}}mulo}, {Carciofi}, {Vieira},
  {Rivinius}, {Faes}, {Figueiredo}, {Bjorkman}, {Georgy}, {Ghoreyshi}, \&
  {Soszy{\'n}ski}}]{rimulo2018}
{R{\'{\i}}mulo}, L.~R., {Carciofi}, A.~C., {Vieira}, R.~G., {et~al.} 2018,
  \mnras, 476, 3555, \dodoi{10.1093/mnras/sty431}

\bibitem[{{Rosenberg} {et~al.}(1975){Rosenberg}, {Eyles}, {Skinner}, \&
  {Willmore}}]{Rosenberg1975}
{Rosenberg}, F.~D., {Eyles}, C.~J., {Skinner}, G.~K., \& {Willmore}, A.~P.
  1975, \nat, 256, 628, \dodoi{10.1038/256628a0}

\bibitem[{{Slettebak}(1982)}]{Slettebak1982}
{Slettebak}, A. 1982, \apjs, 50, 55, \dodoi{10.1086/190820}

\bibitem[{{Smallwood} {et~al.}(2021){Smallwood}, {Martin}, \&
  {Lubow}}]{Smallwood2021}
{Smallwood}, J.~L., {Martin}, R.~G., \& {Lubow}, S.~H. 2021, \apjl, 907, L14,
  \dodoi{10.3847/2041-8213/abd4d6}

\bibitem[{{Smith} {et~al.}(2005){Smith}, {Hazelton}, {Coburn}, {Boggs},
  {Fivian}, {Hurford}, {Hudson}, {Grefenstette}, \& {Gilmore}}]{smith2005}
{Smith}, D.~M., {Hazelton}, B., {Coburn}, W., {et~al.} 2005, The Astronomer's
  Telegram, 557, 1

\bibitem[{{Stella}(1986)}]{Stella1986}
{Stella}, L. 1986, in Plasma Penetration into Magnetospheres, ed. N.~{Kylafis},
  J.~{Papamastorakis}, \& J.~{Ventura}, 199

\bibitem[{{Suffak} {et~al.}(2022){Suffak}, {Jones}, \& {Carciofi}}]{Suffak2022}
{Suffak}, M., {Jones}, C.~E., \& {Carciofi}, A.~C. 2022, \mnras, 509, 931,
  \dodoi{10.1093/mnras/stab3024}

\bibitem[{{von Zeipel}(1910)}]{VonZeipel1910}
{von Zeipel}, H. 1910, Astronomische Nachrichten, 183, 345,
  \dodoi{10.1002/asna.19091832202}

\bibitem[{{Whitlock} {et~al.}(1989){Whitlock}, {Roussel-Dupre}, \&
  {Priedhorsky}}]{Whitlock1989}
{Whitlock}, L., {Roussel-Dupre}, D., \& {Priedhorsky}, W. 1989, \apj, 338, 381,
  \dodoi{10.1086/167206}

\bibitem[{{Wilson} {et~al.}(2002){Wilson}, {Finger}, {Coe}, {Laycock}, \&
  {Fabregat}}]{Wilson2002}
{Wilson}, C.~A., {Finger}, M.~H., {Coe}, M.~J., {Laycock}, S., \& {Fabregat},
  J. 2002, \apj, 570, 287, \dodoi{10.1086/339739}

\end{thebibliography}
\bibliographystyle{aasjournal}



\end{document}